# Exploring the Habitable Zone for *Kepler* planetary candidates

L. Kaltenegger[1,2] and D. Sasselov[2]

[1]MPIA, Koenigstuhl 17, 69117 Heidelberg, Germany
[2]Harvard Smithsonian Center for Astrophysics, 60 Garden St., 02138 MA, Cambridge, USA, email: lkaltene@cfa.harvard.edu

**abstract**
This paper outlines a simple approach to evaluate habitability of terrestrial planets by assuming different types of planetary atmospheres and using corresponding model calculations. Our approach can be applied for current and future candidates provided by the *Kepler* mission and other searches. The resulting uncertainties and changes in the number of planetary candidates in the HZ for the *Kepler* February 2011 data release are discussed. To first order the HZ depends on the effective stellar flux distribution in wavelength and time, the planet albedo, and greenhouse gas effects. We provide a simple set of parameters which can be used for evaluating current and future planet candidates from transit searches.

*Subject headings:* Astrobiology - atmospheric effects - methods: data analysis – Earth - planets and satellites: general – stars: individual (Kepler)

## 1. Introduction

The NASA *Kepler* mission recently announced 1235 planetary candidates (Borucki et al. 2011). We use atmospheric models to explore the potential for habitability of Kepler planetary candidates.

The habitable zone (HZ) concept was proposed for the first time by Huang (1959, 1960) and has been calculated by several authors after that (see e.g. Rasool&DeBergh 1970, Hart 1979,1978, Kasting et al. 1993, Selsis et al. 2007, Williams&Pollard 2002, Peña-Cabrera & Durand-Manterola 2004, Buccino et al. 2006, VonBloh et al. 2007, Spiegel et al. 2009). The main differences are in the climatic constraints imposed on the limits of the HZ by these studies. We focus on the circumstellar HZ, that was defined by Kasting et al. (1993) as an annulus around a star where a planet with an atmosphere and a sufficiently large water content like Earth can host liquid water permanently on a solid surface. This definition of the HZ implies surface habitability because it is defined to allow remote detectability of life as we know it. *Kepler* planetary candidates with radii below 2 Earth's radii are consistent with models of potentially rocky planets.

The two edges of the HZ are influenced by the relationship between the albedo of the planet and the effective temperature of the star (Fig.1). In this definition the inner edge of the HZ is defined as the location where the entire water reservoir can be vaporized by runaway greenhouse conditions, followed by the photo-dissociation of water vapor and subsequent escape of free hydrogen into space. The outer boundary is defined as the distance from the star where the maximum greenhouse effect fails to keep $CO_2$ from condensing permanently, leading to runaway glaciation.

Subsurface life that could exist on planets or moons with very different surface temperatures is not considered, because of the lack of atmospheric features that could be used to remotely assert habitability on such an object (Rosing 2005). We will not discuss the issue of life emerging in other solvents here.

Note that a planet found in the HZ is not necessarily habitable, since many factors may prevent habitability like the lack of water or ingredients necessary for the emergence of life (see e.g. Zahnle et al. 2007, Selsis et al. 2007).

We outline how potential habitability can be evaluated for current and future candidates from the *Kepler* mission and other searches. We introduce the concept of the HZ in §2, discuss the influence of the main parameters in §3 and present our results – the difference in temperature of the recently announced *Kepler* planetary candidates and the location and extent of the HZ for these parameters – in §4. In §5 we discuss important additional uncertainties, and summarize our conclusions in §6.

## 2. The description of the HZ

Different aspects of what determines the boundaries of HZ have been discussed broadly in the literature. Our aim here is to focus on first order effects and how they match the level of information and uncertainties for planets and planetary candidates from transiting searches, and in particular from the NASA *Kepler* mission. Therefore we consider the smallest-size objects in the current *Kepler* sample with radii smaller than 2 Earth radii, because they could possess one of the types of atmospheres we will consider in our study. We assume that, if such planets are rocky, they could possess secondary atmospheres from outgassing that resemble Venus on the hot side and Mars on the cold side of the HZ, and ancient/current Earth near the middle. This range of atmosphere thicknesses and chemistries is sufficiently broad to illustrate the magnitude of the effects on the HZ boundaries

The planets we consider in the definition of the HZ have enough surface water –like Earth - to host liquid water for temperatures between 273K and the critical temperature of water $T_c$ = 647K. Planets with less water will have a narrower HZ. In addition atmospheric $CO_2$ must accumulate in the planet's atmosphere when the mean surface temperature falls below 273K. This is provided on a geologically active planet like Earth, that continuously outgasses $CO_2$ and forms carbonates in the presence of surface water, by the carbonate-silicate cycle, which also stabilizes the long-term climate and atmospheric $CO_2$ content (Walker et al. 1981). The width and distance of HZ annulus – as well as the temperature - for an Earth-like planet depend to a first approximation on 4 main parameters: 1) incident stellar flux which depends on stellar luminosity, spectral energy distribution and eccentricity of the system, 2) planetary albedo, 3) greenhouse gas concentration, and 4) energy distribution in the planetary atmosphere.

## 3. Influence of the Main Parameters on the HZ

The simpler calculation of the equilibrium temperature $T_{eq}$ is often used to determine habitable candidates. It is based on the assumption that the planet radiates as a grey body, i.e., it does not have an atmosphere (see e.g. Borucki et al. 2011). The stellar flux absorbed and radiated by the planet are given in eq. (1) and (2) respectively.

$$F = (1-A)\, \pi r_{pl}^2\, \sigma T_{star}^4\, 4\pi r_{star}^2 / (4\pi D^2) \quad (1)$$

$$f = \beta\, 4\pi\, r_{pl}^2\, \sigma T_{eq\_pl}^4 \quad (2)$$

where $\sigma$ is the Stefan-Boltzmann constant, $r$ the radius, $T$ the temperature, $A$ is the Bond albedo of the planet, $D$ the planet's semi major axis, and $\beta$ represents the fraction of the planetary surface that reradiates the absorbed flux. The parameter $\beta$ has extremes of 1 to 0.5, where for 1 the incident energy is uniformly reradiated by the entire surface of the planet (e.g. for a rapidly rotating planet with an atmosphere, like Earth), and 0.5, where only half of the planet surface is involved in the reradiation, as for a tidally locked, synchronous rotating planet without an atmosphere. One could use a generic emissivity factor in the divisor of eq.(1) to mimic atmospheric warming. We include the greenhouse effect in the atmosphere directly in our calculations (see eq.(6) and Fig.1). By setting f = F, the equilibrium temperature $T_{eq\_pl}$ of a planet is given by:

$$T_{eq\_pl} = T_{star}\, ((1-A)\, r_{star}^2 / 4\beta D^2)^{1/4} \quad (3)$$

The uncertainties in deriving $dT_{eq\_pl}$ (called $dT_{pl}$ hereafter) are discussed in §5. Note that for a planet with a dense atmosphere, like Earth, $T_{eq}$ does not indicate any physical temperature at the surface or in the atmosphere. $T_{eq}$ has to be below 270K for the planet to be habitable. As discussed in detail in Selsis et al. (2007), if the surface temperature of a habitable planet stays below the $T_c$ of water, the thermal emission of the planet can not exceed the greenhouse threshold of about 300W $m^{-2}$ what corresponds to a Black Body radiating at 270K. A planet radiating above 270K either has a $T_{surf} < T_{crit}$, but no liquid water on its surface or a considerable amount of water and a $T_{surf} > 1400K$, what allows the planet to balance the absorbed stellar radiation by radiating in the visible and radio wavelength where water opacity is negligible. Both cases are not habitable for life as we know it. We set the outer limit of the HZ to $T_{eq} < 175K$ here, what corresponds to a planet at 2.4 AU in the present Solar System. Note that these values could be even lower due to selective cloud coverage.

We consider 3 models of Earth-like planets to calculate the edges of the HZ: high $CO_2$, current-Earth or high $H_2O$ atmospheres, for stars with effective temperatures between 3700K and 7200K. These models, representing a planet on the outer, middle and inner edge of the HZ, derive the limits for the HZ, $l_{in}$ and $l_{out}$, given in eq.(4) and eq.(5) (Kasting et al. 1993, Selsis et al. 2007):

$$l_{in} = (l_{in\_sun} - a_{in}T_{star} - b_{in}T_{star}^2)\,(L/L_{sun})^{1/2} \quad (4)$$

$$l_{out} = (l_{out\_sun} - a_{out}T_{star} - b_{out}T_{star}^2)(L/L_{sun})^{1/2} \quad (5)$$

with $a_{in} = 2.7619\ 10^{-5}$, $b_{in} = 3.8095\ 10^{-9}$, $a_{out} = 1.3786\ 10^{-4}$, $b_{out} = 1.4286\ 10^{-9}$, and $T_{Star} = T_{eff} - 5700$, $l_{in}$ and $l_{out}$ (in AU), and $T_{eff}$ (in K).

Depending on the fractional cloud cover, the theoretical outer edge of the HZ for our Sun occurs between 1.67AU in the cloud-free limit, 1.95 AU and 2.4AU for 50% and 100% cloud cover respectively (for more details on cloud effects on the HZ see e.g. Forget&Pierrehumbert 1997, Mischna et al. 2000, vonParis et al. 2008,

Wordsworth et al. 2010, Kitzmann et al. 2010). Note that the water-loss limit for $T_{surf}$ = 373K for the 50% cloud case corresponds to the "Venus water loss limit", empirically derived from Venus position in our Solar System (0.72 AU).

Here we use the limits of the current HZ of our Solar System assuming 0%, 50% and 100% cloud coverage, that gives $l_{in}$ as 0.84-0.95, 0.68-0.76 and 0.46-0.51 and $l_{out}$ as 1.67, 1.95 and 2.4, respectively (see Kasting et al 1993, Selsis et al 2007). The two values quoted for $l_{in}$ represent the runaway greenhouse and the water loss limit ($T_{surf}$=373K), for the inner limit of the HZ. Note that the effect of the spectral type on the albedo (see Fig.1) was estimated for a cloud-free atmosphere, therefore the 100% cloud value is used here to explore the effect of clouds on the HZ (dashed lines).

The $T_{surf}$ sets the surface vapor pressure, especially on the inner edge of the HZ, where most of the Kepler candidates are found (see Fig.2, Fig.3). The water vapor in turn affects $T_{surf}$ because it blocks the outgoing IR radiation, and reduces the atmospheric laps rate. It initially decreases the planet's albedo by absorbing stellar NIR radiation. For $T_{surf}$ > 373K and surface vapor pressure above 1bar, the bond albedo increases, due to the strong Rayleigh backscattering in the visible and the saturation of the NIR water absorption bands. The bond albedo decreases with decrease in stellar temperature because of the redshift of the stellar spectrum that deposits most of the stellar flux onto the planet's atmosphere at wavelength longward of the highly reflective Rayleigh scattering regime (see Fig.1).

## 4. Results

To simply estimate if a planet is potentially habitable (175 < $T_{eq}$ < 270K), one can use eq.(6) to approximate $T_{eq}$ for Earth-like planets around different stars using albedo from Fig.1. Eq.(6) also includes the effect of eccentricity (William & Pollard 2002), Note that for a rapid rotating planet β should be set to one (see discussion).

$$T_{eq}=T_{star}( (1-A)r_{star}^2 / (4\beta D^2(1-e^2)^{1/2}) )^{1/4} \quad (6)$$

Assuming circular orbits and 50% cloud coverage in accordance to the "Venus water loss limit" leads to 27 Kepler planetary candidates with 175K < $T_{eq}$ < 270K. Among those are 3 planetary candidates that have radii smaller than 2 Earth radii (Table 1). The biggest change in $T_{pl}$ results from the change of albedo of the planets depending on the cloud coverage (see Fig.1) what maintains 12 and 2 potentially rocky planets for a clear atmospheres and 67 and 4 potentially rocky planetary candidates for 100% cloud coverage respectively. Eccentricity only has a very small effect on the temperature of the planet, an eccentricity of 0.2 and 0.6, results in an average incident stellar flux increase of 2% and 25% respectively but only a difference of 2 to 10 degrees on $T_{pl}$.

Fig.1 shows the maximum albedo at the inner edge of the HZ (left) and the inner edge of the HZ for water loss (solid lines) and onset greenhouse (dashed lines) for 0%, 50% and 100% cloud coverage for main sequence stars from 3700K to 7200K derived from eq. (4). 1223 of the 1235 stars that host planetary candidates in the Kepler data release have temperatures above 3700K. Using the maximum and minimum albedo values at the inner and outer edge of the HZ respectively and (eq.6), one can calculate a minimum $T_{eq}$ for a planetary candidate to assess habitability. Table 1 and Fig. 2 show the minimum $T_{eq}$ derived for 0%, 50% and 100% cloud coverage using the maximum albedo for Earth-like planets and indicates which rocky planets or satellites could have $T_{eq}$< 270K.

Table 2 and Fig. 3 show the limits of the HZ for 0%, 50% and 100% cloud coverage for all Kepler candidates that could be rocky and potentially be habitable. Note that the HZ is only defined for rocky planets, the focus of our search for habitable conditions.

Applying our analysis to the whole *Kepler* planetary sample of 1235 transiting planetary candidates, assuming the maximum Earth-like Bond albedo at the inner edge of the HZ, results in 12, 27, 67 planetary candidates with $T_{eq}$ smaller than the water evaporation limit for 0%, 50% and 100% clouds, as well as 18, 43, 76 with $T_{eq}$ lower than the runaway greenhouse limit respectively. Among those 2, 3, 6 as well as 3, 4, 6 planets respectively have radii below 2 Earth radii consistent with rocky planets (KOI1026.01, 854.01, 701.03, 268.01, 326.01, 70.03). Note that 2 planets KOI771.01 and KOI99.01 lie outside the outer edge of the HZ for 0% and KOI771.01 outside even for 100% cloud coverage. Many of the 54 planetary candidates announced in the Kepler sample as in the HZ based on simple calculations are with more detailed study (table 2) outside the HZ of their host stars because $T_{eq}$ is above 270K.

Fig. 3 shows that the individual stellar parameters need to be taken into account to determine the limits of the Habitable Zone. For several Kepler planetary candidates the nominal (line) and individual (crosses) stellar limits of the HZ differ (solid lines for 0%, 50% and dashed line for 100% cloud coverage) (see eq.(4) and (5)). In the nominal calculations the radius of the host star is generally derived from stellar models for a stellar temperature. Using real stellar parameters changes the stellar luminosity and therefore $T_{pl}$ and the limits of the HZ (crosses) as seen in Fig.3, e.g. for KOI 701.03 and KOI 268.01.

The location of the Kepler planets is consistent with our expectations to find hot small planets first and cooler ones when further *Kepler* data and thus planets in larger orbits, are available.

## 5. Discussion

The biggest uncertainty in $dT_{pl}$ is due to uncertainties in the stellar parameters (Brown et al. 2011; Borucki et al. 2011). For the Kepler sample these values are on average $dT_{star}/T_{star}$ = 3.6%, $dr_{star}/r_{star}$ = 25%, distance D (assuming negligible uncertainty in the orbital periods) is given by $dD \approx dM_{star}/3M_{star}$ where $dM_{star}$, is about 25%. Using $dT_{pl}/T_{pl} = dT_{star}/T_{star}$, $dT_{pl}/T_{pl} = dr_{star}/2r_{star}$, and $dT_{pl}/T_{pl} = dD/2D$ results in 3.6%, 12.5% and 4% respectively. These errors influence the limits of the HZ substantially as seen in Fig.3 and could explain the difference between some of the nominal and individual stellar limits of the derived HZ. The uncertainty of $dT_{pl}/T_{pl}$ based on planetary parameters is 12.5% due to uncertainties in β (here assumed to be 0.75 +/- 0.25) and a maximum of 12 % due to the range of Bond albedos for the water loss limit and 18% for the greenhouse onset limit (see Fig.1) using $dT_{pl}/T_{pl} = d\beta/4\beta$ and $dT_{pl}/T_{pl} = dA/4(1-A)$. This leads to an accumulated error of 20% to 24% in the estimate of the $T_{pl}$ for the water loss limit and greenhouse onset respectivley (consistent with the 22% Borucki et al. 2011).

Table 2 shows the complete sample of planetary candidates that could be in the HZ assuming the error reduces or increases $T_{pl}$ by 22%. Assuming the errors decreases $T_{eq}$ the number of planetary candidates in the water loss HZ increases the number of planetary candidates from 12, 27, 67 to 45, 67, 124 while it decreases to 4, 5, 19 assuming the error increases $T_{pl}$ by 22% for 0%, 50% and 100% clouds respectively. The number of small, potentially rocky planets in that sample (Table 1) changes from 2, 3 and 6 to 3, 4, and 14 planetary candidates assuming the error reduces $T_{pl}$ (KOI 1026.01, 854.01, 701.03, 268.01, 326.01, 70.03, 314.02, 518.02, 494.01, 1263.01, 899.03, 504.01, 446.02, 1281.01) and to 0, 0, 3 for an increase in $T_{pl}$ by 22% respectively (see Table 1 and 2) for 0%, 50% and 100% clouds respectively.

The HZ does not apply to gas planets, but would apply to rocky satellites that could potentially be detected in the *Kepler* data (see e.g. Kaltenegger 2010, Kipping et al. 2010). A detailed discussion of limits of habitability of exomoons is presented in Williams et al. (1997). Using the most extreme limits for the inner edge of the HZ – empirically determined by the current flux at Venus' orbit, what is about 30% higher than the initial flux at Venus' orbit – would result in 41 planets in the HZ with $T_{eq}$ between 175K and 307K, four of those have radii consistent with a rocky planet (KOI701.03, 1026.01, 268.01 and 854.01).

Due to the close orbit of the planetary candidates to their star, tidal locking can be explored using the reradiation parameter β. However detailed models for Earth have shown that even for planets in synchronous rotation direct illumination of only one hemisphere does not prevent habitability for planets with even modestly dense atmospheres (Haberle et al. 1996, Joshi et al. 1997, Joshi 2003, Edson et al. 2011, Wordsworth et al. 2010, Heng et al. 2011), provided atmospheric cycles transport heat from the dayside to the nightside. Therefore the limits of the HZ are unlikely to change significantly due to synchronous rotation for Earth-like planets.

## 6. Conclusions

The NASA *Kepler* mission recently announced 1235 planetary candidates. We use atmospheric models to explore the potential for habitability of Kepler planetary candidates. Table 1-2 and Fig.2-3 show the temperature as well as the extent of the HZ for the Kepler planetary candidates that could potentially be habitable. Many of the 54 planetary candidates announced in the Kepler sample in the HZ, based on simple calculations, are - with a more detailed study, outside the HZ of their host stars, because $T_{eq}$ is above 270K (see Table 1 for all potentially rocky planets).

The three main points of this paper are: (1) to provide a simple approximation (eq.6) for habitability of Earth-like planets from the data provided by transit surveys like Kepler (using maximum albedo values from atmospheric models Fig.1) to assess their potential as a habitat, (2) to demonstrate that one needs to consider the individual stellar parameters to determine the limits of the HZ, and (3) to explore the change in the sample located in the HZ due to individual factors in eq.(6) and associated errors.

Applying our analysis to the whole *Kepler* planetary sample of 1235 transiting planetary candidates, assuming the maximum Earth-like Bond albedo for rocky planet atmospheres (see Fig.1, Table 2), results in 12, 27, 67 planetary candidates with $T_{eq}$ smaller than the water loss limit ($T_{surf}$ = 373K) for 0%, 50% and 100% clouds respectively, and 18, 43, 76 planetary candidates with temperatures lower than the runaway greenhouse limit respectively. Among those are 2, 3, 6 as well as 3, 4, 6 planets respectively, that have radii below 2 Earth radii consistent with rocky planets (KOI1026.01, 854.01, 701.03, 268.01, 326.01, 70.03).

The potentially rocky planet candidates in multiple systems in the *Kepler* February 2011 data release, KOI701.3 and KOI70.3 are extremely interesting objects because their mass could be determined using transit time variations to calculate a mean density and potentially confirm high density and rocky characteristics (Table 1). Assuming errors will reduce $T_{eq}$ (see Discussion) KOI314.02, 899.03, 446.02, 518.02, and 70.03 are also part of that sample.

Note that even if these small planetary candidates can be confirmed to have a mean density consistent with a rocky composition, many aspects can prevent a planet in the HZ from being habitable, e.g. limited amount of water or other ingredients essential for life. Therefore the atmosphere of planets has to be characterized to explore if the planet is a potential habitat or shows signs of life.

Due to the large average distance of 500-1000pc to its target stars, *Kepler*'s results can only provide statistics of the amount of planets per star. That, as well as the increasing number of small potentially rocky planets shown by the new *Kepler* results, strengthens the scientific case for a mission to find and characterize such small planets orbiting stars close to our Sun.

**Acknowledgement**
We are grateful to Jim Kasting and Franck Selsis for comments and discussion. L.K. acknowledges support from NAI and DFG funding ENP Ka 3142/1-1.

**Table 1** shows the characteristics of the potentially rocky habitable *Kepler* planetary candidates (KOI, $R_{pl}$, Period, $T_{eff}$, $R_{star}$, $T_{eq}$) for 0, 50% and 100% clouds.

| KOI | $R_{pl}$ | D AU | $T_{eff}$ (K) | $R_{star}$ | $T_{eq}$ Kepler | $T_{eq}$ (K) Water loss | | | $T_{eq}$ (K) runaway GH | | |
|---|---|---|---|---|---|---|---|---|---|---|---|
| | | | | | | 0% | 50% | 100% | 0% | 50% | 100% |
| 1026.01 | 1.8 | 0.325 | 3802 | 0.68 | 243 | 256 | 230 | 191 | 241 | 218 | 182 |
| 854.01 | 1.9 | 0.217 | 3743 | 0.49 | 248 | 262 | 235 | 195 | 247 | 223 | 186 |
| 701.03 | 1.7 | 0.454 | 4869 | 0.68 | 263 | 275 | 247 | 203 | 259 | 234 | 194 |
| 268.01 | 1.8 | 0.406 | 4808 | 0.79 | 296 | 310 | 278 | 229 | 292 | 263 | 218 |
| 326.01 | 0.9 | 0.05 | 3240 | 0.27 | 332 | 352 | 316 | 263 | 332 | 300 | 250 |
| 70.03 | 2 | 0.35 | 5342 | 0.7 | 333 | 347 | 310 | 255 | 326 | 294 | 243 |
| 314.02 | 1.6 | 0.128 | 3900 | 0.61 | 375 | 396 | 356 | 295 | 373 | 338 | 281 |
| 518.02 | 1.9 | 0.23 | 4822 | 0.76 | 387 | 405 | 363 | 299 | 381 | 344 | 285 |
| 494.01 | 1.8 | 0.157 | 4854 | 0.52 | 390 | 408 | 366 | 302 | 384 | 346 | 287 |
| 1263.01 | 2 | 0.182 | 4007 | 0.9 | 393 | 414 | 372 | 308 | 390 | 353 | 294 |
| 899.03 | 1.7 | 0.091 | 3653 | 0.55 | 396 | 418 | 376 | 312 | 395 | 357 | 298 |
| 504.01 | 1.7 | 0.228 | 5403 | 0.68 | 412 | 428 | 383 | 315 | 402 | 363 | 299 |
| 446.02 | 1.7 | 0.164 | 4492 | 0.7 | 409 | 430 | 386 | 319 | 405 | 366 | 304 |
| 1281.01 | 2 | 0.265 | 5546 | 0.82 | 430 | 446 | 400 | 328 | 420 | 378 | 312 |
| 1591.01 | 1.5 | 0.134 | 5130 | 0.49 | 433 | 451 | 404 | 333 | 425 | 383 | 317 |
| 148.03 | 2 | 0.235 | 5063 | 0.89 | 435 | 454 | 407 | 335 | 427 | 385 | 319 |
| 663.02 | 1.7 | 0.124 | 4156 | 0.7 | 435 | 459 | 412 | 341 | 432 | 391 | 325 |
| 718.03 | 1.5 | 0.261 | 5801 | 0.78 | 442 | 457 | 409 | 335 | 430 | 387 | 318 |
| 560.01 | 1.8 | 0.154 | 5142 | 0.59 | 444 | 463 | 415 | 342 | 436 | 393 | 325 |
| 486.01 | 1.4 | 0.152 | 5625 | 0.51 | 454 | 471 | 421 | 346 | 443 | 399 | 328 |
| 1435.01 | 1.7 | 0.234 | 5744 | 0.75 | 454 | 469 | 420 | 344 | 441 | 397 | 327 |

**Table 2** shows the two inner limits and outer limit of the HZ for *Kepler* planetary candidates in the HZ of their stars for a 0%, 50% and 100% clouds, sorted by $T_{eq}$.

| KOI | $R_{pl}$ | D | $T_{eff}$ | $R_{star}$ | $HZ_{in}$ cloud: 0% | $HZ_{out}$ | $HZ_{in}$ 50% | $HZ_{out}$ | $HZ_{in}$ 100% | $HZ_{out}$ |
|---|---|---|---|---|---|---|---|---|---|---|---|
| 99.01 | 4.6 | 1.693 | 4951 | 1.11 | 0.70 | 0.79 | 1.46 | 0.57 | 0.64 | 1.69 | 0.39 | 0.43 | 2.05 |
| 1439.01 | 4.1 | 2.235 | 5967 | 1.22 | 1.09 | 1.23 | 2.14 | 0.88 | 0.98 | 2.51 | 0.59 | 0.66 | 3.10 |
| 1477.01 | 9.4 | 1.044 | 5346 | 0.71 | 0.52 | 0.59 | 1.05 | 0.42 | 0.47 | 1.22 | 0.29 | 0.32 | 1.50 |
| 868.01 | 12.5 | 0.629 | 4118 | 0.71 | 0.32 | 0.36 | 0.68 | 0.26 | 0.29 | 0.79 | 0.18 | 0.20 | 0.95 |
| 881.02 | 3.9 | 0.693 | 5053 | 0.6 | 0.39 | 0.45 | 0.81 | 0.32 | 0.36 | 0.94 | 0.22 | 0.24 | 1.15 |
| 683.01 | 4.2 | 0.839 | 5624 | 0.78 | 0.63 | 0.71 | 1.25 | 0.51 | 0.57 | 1.46 | 0.34 | 0.38 | 1.79 |
| 1582.01 | 4.5 | 0.626 | 5384 | 0.64 | 0.47 | 0.53 | 0.96 | 0.38 | 0.43 | 1.12 | 0.26 | 0.29 | 1.37 |
| 1503.01 | 2.7 | 0.535 | 5356 | 0.56 | 0.41 | 0.46 | 0.83 | 0.33 | 0.37 | 0.97 | 0.23 | 0.25 | 1.19 |
| 1099.01 | 3.7 | 0.573 | 5665 | 0.55 | 0.45 | 0.51 | 0.89 | 0.36 | 0.40 | 1.04 | 0.25 | 0.27 | 1.28 |
| 433.02 | 13.4 | 0.935 | 5237 | 1.08 | 0.76 | 0.86 | 1.55 | 0.62 | 0.69 | 1.80 | 0.42 | 0.47 | 2.20 |
| 1026.01 | 1.8 | 0.325 | 3802 | 0.68 | 0.26 | 0.29 | 0.57 | 0.21 | 0.24 | 0.65 | 0.15 | 0.16 | 0.79 |
| 1486.01 | 8.5 | 0.796 | 5688 | 0.83 | 0.68 | 0.77 | 1.36 | 0.55 | 0.61 | 1.58 | 0.37 | 0.41 | 1.94 |
| 854.01 | 1.9 | 0.217 | 3743 | 0.49 | 0.18 | 0.20 | 0.40 | 0.15 | 0.17 | 0.46 | 0.10 | 0.11 | 0.55 |
| 351.01 | 8.5 | 0.966 | 6103 | 0.94 | 0.87 | 0.99 | 1.71 | 0.70 | 0.79 | 2.00 | 0.47 | 0.53 | 2.48 |
| 465.01 | 4.8 | 1.002 | 6029 | 1.00 | 0.91 | 1.03 | 1.78 | 0.73 | 0.82 | 2.09 | 0.49 | 0.55 | 2.58 |
| 701.03 | 1.7 | 0.454 | 4869 | 0.68 | 0.42 | 0.47 | 0.87 | 0.34 | 0.38 | 1.00 | 0.23 | 0.26 | 1.22 |
| 211.01 | 9.6 | 1.048 | 6072 | 1.09 | 1.00 | 1.14 | 1.97 | 0.81 | 0.91 | 2.30 | 0.54 | 0.61 | 2.85 |
| 113.01 | 8.3 | 0.888 | 5362 | 1.15 | 0.84 | 0.95 | 1.71 | 0.69 | 0.77 | 1.99 | 0.47 | 0.52 | 2.44 |
| 1429.01 | 4.2 | 0.69 | 5595 | 0.87 | 0.69 | 0.78 | 1.39 | 0.56 | 0.63 | 1.61 | 0.38 | 0.42 | 1.98 |
| 1423.01 | 4.3 | 0.475 | 5288 | 0.66 | 0.47 | 0.53 | 0.96 | 0.38 | 0.43 | 1.12 | 0.26 | 0.29 | 1.37 |
| 902.01 | 5.7 | 0.324 | 4312 | 0.65 | 0.32 | 0.36 | 0.68 | 0.26 | 0.29 | 0.78 | 0.18 | 0.20 | 0.94 |
| 1008.01 | 204.8 | 0.895 | 6228 | 0.95 | 0.91 | 1.04 | 1.78 | 0.74 | 0.83 | 2.09 | 0.49 | 0.55 | 2.59 |
| 87.01 | 2.4 | 0.877 | 5606 | 1.14 | 0.91 | 1.03 | 1.82 | 0.74 | 0.82 | 2.12 | 0.50 | 0.55 | 2.61 |
| 1168.01 | 3.9 | 1.221 | 6209 | 1.37 | 1.31 | 1.49 | 2.55 | 1.06 | 1.19 | 3.00 | 0.71 | 0.79 | 3.71 |
| 139.01 | 5.7 | 0.741 | 5921 | 0.9 | 0.79 | 0.90 | 1.56 | 0.64 | 0.72 | 1.83 | 0.43 | 0.48 | 2.25 |
| 1134.02 | 9.7 | 0.646 | 4904 | 1.1 | 0.68 | 0.77 | 1.42 | 0.56 | 0.62 | 1.64 | 0.38 | 0.42 | 2.00 |
| 1361.01 | 2.2 | 0.243 | 4050 | 0.59 | 0.25 | 0.29 | 0.55 | 0.21 | 0.23 | 0.63 | 0.14 | 0.16 | 0.77 |
| 1232.01 | 48.6 | 0.75 | 2191 | 6.16 | 0.79 | 0.89 | 1.91 | 0.65 | 0.72 | 2.15 | 0.45 | 0.50 | 2.55 |
| 1208.01 | 7 | 0.913 | 6293 | 1.07 | 1.05 | 1.19 | 2.03 | 0.85 | 0.95 | 2.39 | 0.57 | 0.63 | 2.96 |
| 1375.01 | 17.9 | 0.958 | 6169 | 1.17 | 1.11 | 1.25 | 2.16 | 0.89 | 1.00 | 2.53 | 0.60 | 0.67 | 3.14 |
| 536.01 | 3 | 0.588 | 5614 | 0.84 | 0.67 | 0.76 | 1.35 | 0.54 | 0.61 | 1.57 | 0.37 | 0.41 | 1.93 |
| 1472.01 | 3.6 | 0.37 | 5455 | 0.56 | 0.42 | 0.48 | 0.86 | 0.34 | 0.38 | 1.00 | 0.23 | 0.26 | 1.22 |
| 806.01 | 9 | 0.53 | 5206 | 0.88 | 0.61 | 0.69 | 1.25 | 0.50 | 0.55 | 1.45 | 0.34 | 0.38 | 1.77 |
| 375.01 | 8.8 | 0.729 | 5692 | 1.04 | 0.85 | 0.96 | 1.70 | 0.69 | 0.77 | 1.98 | 0.47 | 0.52 | 2.44 |
| 268.01 | 1.8 | 0.406 | 4808 | 0.79 | 0.47 | 0.53 | 0.99 | 0.39 | 0.43 | 1.14 | 0.26 | 0.29 | 1.39 |
| 351.02 | 6 | 0.713 | 6103 | 0.94 | 0.87 | 0.99 | 1.71 | 0.70 | 0.79 | 2.00 | 0.47 | 0.53 | 2.48 |
| 865.01 | 6 | 0.473 | 5560 | 0.73 | 0.57 | 0.65 | 1.15 | 0.46 | 0.52 | 1.34 | 0.32 | 0.35 | 1.65 |
| 1226.01 | 26.6 | 0.514 | 5045 | 0.94 | 0.62 | 0.70 | 1.27 | 0.50 | 0.56 | 1.47 | 0.34 | 0.38 | 1.80 |
| 1032.01 | 26.2 | 1.558 | 4787 | 3.15 | 1.87 | 2.11 | 3.91 | 1.53 | 1.70 | 4.52 | 1.05 | 1.16 | 5.49 |
| 1463.01 | 16.3 | 0.795 | 6020 | 1.09 | 0.99 | 1.12 | 1.94 | 0.80 | 0.89 | 2.27 | 0.54 | 0.60 | 2.81 |
| 998.01 | 25 | 0.592 | 5814 | 0.86 | 0.73 | 0.83 | 1.45 | 0.59 | 0.66 | 1.70 | 0.40 | 0.44 | 2.09 |
| 682.01 | 4.9 | 0.591 | 5504 | 0.95 | 0.73 | 0.83 | 1.48 | 0.59 | 0.66 | 1.72 | 0.40 | 0.45 | 2.11 |

| | | | | | | | | | | | | |
|---|---|---|---|---|---|---|---|---|---|---|---|---|
| 959.01 | 5.3 | 0.063 | 1644 | 1.05 | 0.08 | 0.09 | 0.19 | 0.06 | 0.07 | 0.21 | 0.04 | 0.05 | 0.25 |
| 812.03 | 2.1 | 0.206 | 4097 | 0.57 | 0.25 | 0.28 | 0.55 | 0.21 | 0.23 | 0.63 | 0.14 | 0.16 | 0.75 |
| 364.01 | 2.6 | 0.619 | 5551 | 1.01 | 0.79 | 0.89 | 1.59 | 0.64 | 0.72 | 1.85 | 0.44 | 0.48 | 2.27 |
| 416.02 | 2.8 | 0.376 | 5083 | 0.75 | 0.50 | 0.56 | 1.03 | 0.41 | 0.45 | 1.19 | 0.28 | 0.31 | 1.45 |
| 1596.02 | 3.5 | 0.416 | 4656 | 0.99 | 0.56 | 0.63 | 1.17 | 0.46 | 0.51 | 1.35 | 0.31 | 0.35 | 1.64 |
| 51.01 | 4.8 | 0.056 | 3240 | 0.27 | 0.08 | 0.08 | 0.17 | 0.06 | 0.07 | 0.19 | 0.04 | 0.05 | 0.23 |
| 422.01 | 16.5 | 0.692 | 6002 | 1.09 | 0.98 | 1.11 | 1.93 | 0.79 | 0.89 | 2.26 | 0.53 | 0.59 | 2.79 |
| 1574.01 | 5.8 | 0.465 | 5537 | 0.85 | 0.66 | 0.75 | 1.33 | 0.54 | 0.60 | 1.55 | 0.36 | 0.40 | 1.90 |
| 622.01 | 9.3 | 0.568 | 5171 | 1.17 | 0.80 | 0.91 | 1.65 | 0.65 | 0.73 | 1.91 | 0.45 | 0.49 | 2.33 |
| 1268.01 | 8.6 | 0.671 | 6064 | 1.06 | 0.97 | 1.10 | 1.91 | 0.79 | 0.88 | 2.24 | 0.53 | 0.59 | 2.76 |
| 1261.01 | 6.3 | 0.521 | 5760 | 0.9 | 0.75 | 0.85 | 1.50 | 0.61 | 0.68 | 1.75 | 0.41 | 0.46 | 2.15 |
| 555.02 | 2.3 | 0.376 | 5218 | 0.78 | 0.54 | 0.61 | 1.11 | 0.44 | 0.49 | 1.29 | 0.30 | 0.33 | 1.58 |
| 70.03 | 2 | 0.35 | 5342 | 0.7 | 0.51 | 0.58 | 1.04 | 0.41 | 0.46 | 1.21 | 0.28 | 0.31 | 1.48 |
| 1454.01 | 9.2 | 0.487 | 4015 | 1.66 | 0.71 | 0.79 | 1.53 | 0.58 | 0.64 | 1.76 | 0.40 | 0.44 | 2.12 |
| 1527.01 | 4.9 | 0.67 | 5470 | 1.31 | 1.00 | 1.13 | 2.01 | 0.81 | 0.90 | 2.34 | 0.55 | 0.61 | 2.87 |
| 1328.01 | 4.8 | 0.362 | 5425 | 0.72 | 0.54 | 0.61 | 1.09 | 0.44 | 0.49 | 1.27 | 0.30 | 0.33 | 1.56 |
| 564.02 | 5 | 0.505 | 5686 | 0.93 | 0.76 | 0.86 | 1.52 | 0.62 | 0.69 | 1.77 | 0.42 | 0.46 | 2.18 |
| 1355.01 | 2.8 | 0.266 | 5529 | 0.52 | 0.40 | 0.46 | 0.81 | 0.33 | 0.37 | 0.95 | 0.22 | 0.25 | 1.16 |
| 1478.01 | 3.7 | 0.348 | 5441 | 0.7 | 0.53 | 0.60 | 1.07 | 0.43 | 0.48 | 1.24 | 0.29 | 0.32 | 1.52 |
| 372.01 | 8.5 | 0.499 | 5638 | 0.95 | 0.77 | 0.86 | 1.53 | 0.62 | 0.69 | 1.79 | 0.42 | 0.47 | 2.19 |
| 711.03 | 2.6 | 0.494 | 5488 | 1.00 | 0.77 | 0.87 | 1.55 | 0.62 | 0.69 | 1.80 | 0.42 | 0.47 | 2.21 |
| 326.01 | 0.9 | 0.05 | 3240 | 0.27 | 0.08 | 0.08 | 0.17 | 0.06 | 0.07 | 0.19 | 0.04 | 0.05 | 0.23 |
| 1426.03 | 36.1 | 0.568 | 5854 | 1.05 | 0.91 | 1.02 | 1.79 | 0.73 | 0.82 | 2.10 | 0.49 | 0.55 | 2.58 |
| 415.01 | 7.7 | 0.611 | 5823 | 1.15 | 0.98 | 1.11 | 1.95 | 0.79 | 0.89 | 2.28 | 0.54 | 0.60 | 2.80 |
| 448.02 | 3.8 | 0.21 | 4264 | 0.71 | 0.34 | 0.38 | 0.73 | 0.28 | 0.31 | 0.84 | 0.19 | 0.21 | 1.01 |
| 1564.01 | 3.1 | 0.275 | 5709 | 0.56 | 0.46 | 0.52 | 0.92 | 0.37 | 0.42 | 1.07 | 0.25 | 0.28 | 1.32 |
| 157.05 | 3.2 | 0.481 | 5675 | 1.00 | 0.82 | 0.92 | 1.63 | 0.66 | 0.74 | 1.90 | 0.45 | 0.50 | 2.33 |
| 401.02 | 6.6 | 0.591 | 5264 | 1.4 | 0.99 | 1.12 | 2.03 | 0.81 | 0.90 | 2.35 | 0.55 | 0.61 | 2.88 |
| 374.01 | 3.3 | 0.628 | 5829 | 1.26 | 1.08 | 1.22 | 2.14 | 0.87 | 0.98 | 2.50 | 0.59 | 0.65 | 3.08 |
| 1192.01 | 4.1 | 0.491 | 5399 | 1.13 | 0.84 | 0.95 | 1.70 | 0.68 | 0.76 | 1.98 | 0.46 | 0.51 | 2.42 |
| 174.01 | 2.5 | 0.267 | 4654 | 0.8 | 0.45 | 0.51 | 0.95 | 0.37 | 0.41 | 1.09 | 0.25 | 0.28 | 1.33 |
| 365.01 | 2.3 | 0.368 | 5389 | 0.86 | 0.64 | 0.72 | 1.29 | 0.52 | 0.58 | 1.50 | 0.35 | 0.39 | 1.84 |
| 947.01 | 2.7 | 0.146 | 3829 | 0.64 | 0.25 | 0.28 | 0.54 | 0.20 | 0.23 | 0.62 | 0.14 | 0.15 | 0.75 |
| 1162.01 | 3.9 | 0.594 | 5833 | 1.26 | 1.08 | 1.22 | 2.14 | 0.87 | 0.98 | 2.50 | 0.59 | 0.65 | 3.08 |

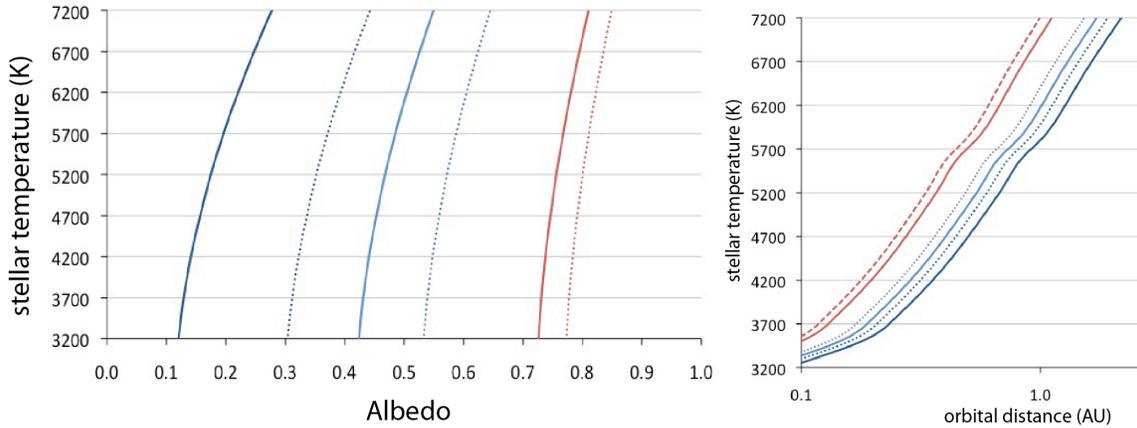

**Fig.1** (left) Maximum albedo at the inner edge of the HZ and (right) the inner edge of the HZ for water loss (solid lines) and onset greenhouse (dashed lines) for 0%, 50% and 100% cloud coverage.

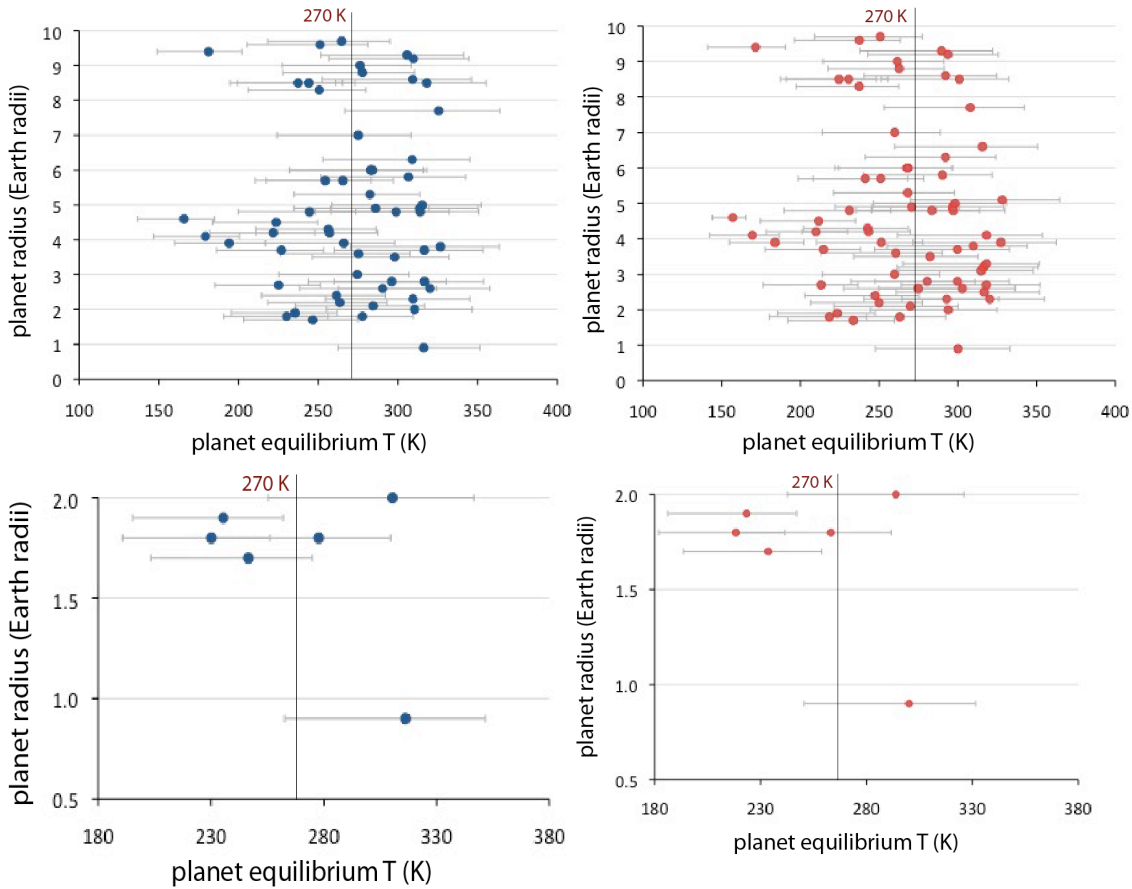

**Fig Fig. 2** Minimum equilibrium Temperature of the Kepler planets candidates (left) water loss limit and onset greenhouse (right) for 50% cloud coverage. Error bars indicate 0% and 100% cloud cover, the line indicates 270K. Detail for the 6 pot. rocky planets (lower pannel).

Kaltenegger & Sasselov 2011                                       accepted ApJL

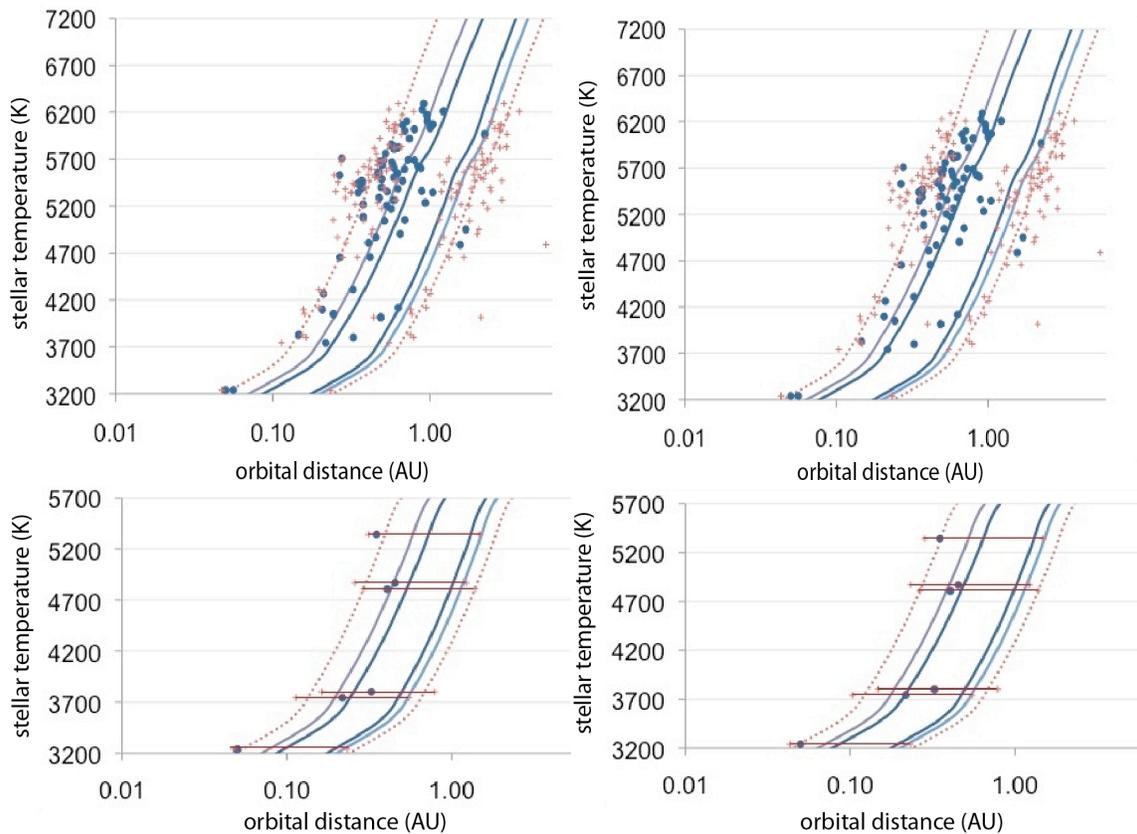

**Fig. 3** Extent of the HZ for (left) water loss limit for 0% and 50% cloud coverage (inner limits) and 100% cloud coverage (outer limit dashed line), (right) runaway greenhouse onset and position of potentially habitable *Kepler* planetary candidates in the HZ, individual HZ limits are indicated with crosses . Detail for the 6 potentially rocky planets (lower pannel).